\documentclass[doublecol]{epl2} 
\newcommand{\beq }{\begin{eqnarray}}
\newcommand{\eeq }{\end{eqnarray}}
\title{Energy Dissipation and Fluctuation-Response 
in Driven Quantum Langevin Dynamics}
\shorttitle{Energy Dissipation and Fluctuation-Response 
in Driven Quantum Langevin dynamics}

\author{Keiji Saito\inst{1,2}}
\shortauthor{Keiji Saito}

\institute{                    
  \inst{1} Graduate School of Science, University of Tokyo, 113-0033,
  Japan \\
\inst{2} CREST, Japan Science and Technology (JST), Saitama, 332-0012, Japan
}

\abstract{
Energy dissipation in a nonequilibrium steady state is studied in driven 
quantum Langevin systems. We study energy dissipation flow to thermal 
environment, and obtain a general formula for the 
average rate of energy dissipation using
an autocorrelation function for the system variable. This leads to
a general expression of the equality that connects 
the violation of the fluctuation-response relation to the rate of 
energy dissipation, the classical version of which was 
first studied by Harada and Sasa. We also point out that 
the expression depends on coupling form between system and reservoir.}
\begin{document}

\maketitle

\section{Introduction}
Recent developments in nonequilibrium statistical mechanics have
clarified fundamental aspects of nonequilibrium 
fluctuations of work, power flux, heat absorbed etc 
\cite{evan1,lebo99,kur98,jarz97,crooks99,zon04}. 
The fluctuation theorem (FT) is one of the most remarkable discoveries 
in nonequilibrium statistical mechanics.
This theorem quantifies the probability of negative entropy,
which can be important for short measurement times in small systems,
and provides a precise statement of the second law of thermodynamics.
The relation between the transient version of FT and 
the Jarzynski equality has been demonstrated and clarified \cite{crooks99}. 
Both the FT and the Jarzynski relation \cite{jarz97} 
have been tested 
experimentally in systems, 
such as micromechanically manipulated biomolecules
\cite{liphardt02,collin05}, 
colloids in time-dependent laser traps\cite{wang,blickle06},
and optically driven single two-level systems\cite{schuler05}. 
Studying robust properties valid in far-from-equilibrium regime
is obviously important for understanding the general structures of
nonequilibrium statistical mechanics.

Another important aspect of fluctuations is 
fluctuation-response.
In the linear response regime, 
a fluctuation-dissipation relation (FDR) relates a
response function with an 
auto correlation function for physical quantities at equilibrium
\cite{einstein,nyquist,kubo}. However, the FDR is generically 
violated if the system is driven into a nonequilibrium state beyond 
this regime.
Several recent studies considered 
extensions of FDR to far-from-equilibrium regime\cite{harada_sasa2005,speck_seifert,blickle}. 
In Refs.\cite{speck_seifert,blickle},
the {\em violation function} was introduced 
to generalize the Einstein relation, and 
its validity was experimentally studied.
Harada and Sasa considered the relationship between 
the degree of violation of FDR and the rate of 
energy dissipation in over-damped Langevin dynamics, and 
found {\em an equality valid in a far-from-equilibrium regime} 
\cite{harada_sasa2005}. 
The equality was first derived for the following
\beq
\gamma \dot{x} (t) = F(x(t) , t ) + \eta (t) + \varepsilon f(t) ,
\eeq
where $x$ are the coordinates, $\eta$ is the white Gaussian noise 
satisfying $\langle \eta(t) \eta( u) \rangle = 
2 \gamma k_{\rm B} T \delta (t -u )$, and $F(x(t),t)$ represents
a force dependent on space and time. Let $C(t)$ be the autocorrelation 
function for the velocity fluctuation without perturbation, $\varepsilon=0$
\beq
C(t)
 &=& \langle [ \dot{x} (t) - v_s ] [ \dot{x} (0) - v_s ] \rangle  ,
\eeq
where $\langle ...\rangle$ denotes an ensemble average over thermal noise,
and $v_s$ is the average velocity of the particle.
When the external perturbation is finite $\varepsilon \ne 0$,
the response of the velocity obeys the linear response form \cite{kubo}
\beq
\langle \delta \dot{x} \rangle 
&=& \varepsilon 
\int_{-\infty }^{\infty } dt' \chi (t , u ) f (u ) .
\label{deviation_of_v}
\eeq
Then the following equality was derived in Ref.\cite{harada_sasa2005} 
\beq
I &=& \gamma \Bigl\{ v_s^2 
+ \int_{-\infty}^{\infty} {d\omega \over 2\pi } 
\left[ C(\omega ) - 2 k_{\rm B} T \chi  ' (\omega ) \right]
\Bigr\} \label{HS} ,
\eeq
where $I$ is the rate of energy dissipation.
$C(\omega )$ is the Fourier transformation of $C(t)$
and $\chi '( \omega )$ is the real part of  
the Fourier transformation of $\chi(t, u )$.
In an equilibrium state with no 
dissipation flow $I=0$, the FDR $C(\omega ) = 2 k_{\rm B} T \chi (\omega )$ 
is satisfied. On the other hand, 
in a nonequilibrium state, the degree of violation of FDR is related 
to the rate of energy dissipation.
This equality was generalized to correlated thermal noise by 
Deutsch and Narayan \cite{deutsch_narayan}. Recently an experimental 
test of Eq.(\ref{HS}) was performed in an optically driven colloidal 
system \cite{sano_experiment}. 
In general, it is difficult to conduct direct experimental 
measurements of energy dissipation flow. Eq.(\ref{HS}) 
suggests that the flow can 
be obtained with measurable functions in Langevin dynamics.
Thus Eq.(\ref{HS}) is also of practical importance, providing 
a new protocol for measuring 
energy dissipation flow in driven Langevin dynamics.

In this paper, we consider a {\em driven quantum Langevin dynamics} and 
derive a wider class of relations that includes the quantum version
of Eq.(\ref{HS}).
Quantum Langevin dynamics has a wide variety of
applications in areas such as electronic circuits, 
superconducting tunnel junctions,
and electronic systems in semiconductors. \cite{weiss}.
Many different systems can be mapped onto a simple driven quantum Langevin 
equation.
We derive a general relation, which 
reproduces the Callen-Welton FDR in an equilibrium state \cite{callen_welton}.
Our approach provides a unified method for investigating 
energy dissipation flow. This provides a new consideration 
of quantum energy dissipation with respect to Langevin dynamics, 
even if it includes nonlinear couplings between the system and the 
reservoirs.

\section{Quantum Langevin Equation}
We consider a driven quantum Langevin system described as
\beq
{\cal H} (t) &=& {p^2 \over 2m } + V(x,t)
\nonumber \\
&+& \sum_{\ell} 
\Bigl[
{p_{\ell}^2 \over 2 m_{\ell} }
+ {m_{\ell} \omega_{\ell}^2 \over 2}
\left( x_{\ell} - {\lambda_{\ell} x \over m_{\ell }  \omega_{\ell}^2 
} \right)^2
\Bigr] . \label{ham1}
\eeq
where $\{m,x,p\}$ refer to the system degrees of freedom, and  
$\{x_\ell,p_\ell,m_\ell,\omega_\ell\}$ refers to the reservoir. 
Those variables satisfy the commutation relations 
$\left[ x , x_{\ell} \right] = \left[ x_{\ell} , x_{\ell '}\right]=0,
\left[ x, p \right] = i\hbar$, and 
$\left[ x_{\ell} , p_{\ell '} \right] = i\hbar \delta_{\ell , \ell'}$
The potential term $V(x, t)$ drives the system into 
a nonequilibrium steady state. Detailed form is not provided here. 
The coupling constant between the system and bath oscillators
$\{\lambda_\ell \}$ is switched on at time $t_{ini}=-\infty$. The
initial density matrix is assumed to be of the product form
$\rho_{ini} =\rho_S   \otimes \rho_R$,
where $S$ and $R$ refer to the system and the reservoir, respectively.
These matrices are equilibrium distributions.
The reservoir's density matrix is $\rho_{R} ={e^{-\beta {\cal H}_{R }}}
/{{\rm Tr} [e^{-\beta {\cal H}_{R}}]}$ for 
$\beta = 1/(k_{\rm B} T )$. 
By eliminating the bath's
degrees of freedom, we obtain
a quantum Langevin equation easily\cite{weiss}, which is expressed as
\beq
m{\partial^2 x \over \partial t^2}&=& 
- {\partial V(x,t) \over \partial x}-
\int_{0}^{\infty} d u  \gamma ( u ) \dot{x} (t - u )  
+ \eta (t) , \nonumber 
\\ \label{qle}
\eeq
where $\eta$ and $\gamma (t)$ represent a noise term and 
the memory kernels, respectively. 
These terms control dissipation effects from the bath.
The properties of the noise and dissipation are completely determined
by the initial condition of the bath. We define the spectral function
\beq
J (\omega )={\pi \over 2} \sum_\ell {\lambda_\ell^2 \over m_\ell
  \omega_\ell}\delta (\omega-\omega_\ell) .
\eeq 
Then, the dissipation kernels and noise correlations are given by
\beq
\gamma (t)&=&{2 \over \pi} \int_0^\infty d \omega {J(\omega )
\over \omega }
\cos{\omega t}  , \nonumber \\
\langle \eta ( t ) \eta ( u ) \rangle 
&=& {\hbar \over \pi }
\int_{-\infty}^{\infty} d\omega 
e^{-i\omega ( t- u )}
{\omega \over |\omega |}J( | \omega  | )
(1+f(\omega)) , \nonumber 
\eeq
where $f (\omega )=1/(e^{\beta\hbar\omega }-1)$. 
The Fourier transformation of the memory kernel is defined as 
${\gamma} (\omega ) = \int_0^\infty dt
\gamma (t) e^{i \omega t}$. 
Then, the real part of the Fourier transformation
$\gamma \, ' (\omega )$ is expressed as
\beq
\gamma\,' (\omega ) &=& J(|\omega |) / | \omega | . 
\eeq

\section{Response Function}
In the Langevin dynamics,
we consider the relationship between the 
response function and energy dissipation. 
The response function we consider is defined as the response of $\dot{x}$
against a perturbation $-\varepsilon f(t) x$.
The formal expression of response function $\chi (t,u)$ is calculated from 
the standard linear response derivation \cite{kubo}.
In the first order of $\varepsilon$, we obtain the deviation of density matrix 
from the unperturbed one as
\beq
\delta \rho (t) \!\!\!\! &=&\!\!\!\!
{- \varepsilon \over i \hbar}
\int_{t_{ini}}^{t} d u f(u) U(t, t_{ini} ) 
\left[ x( u ) , \rho (t_{ini} ) \right]
U^{\dagger } (t, t_{ini} ) ,
\nonumber \\
&& 
\eeq
where $\rho (t )$ is the density matrix at time $t$, and 
$U(t,t_{ini})$ is the time-evolution operator defined as
\beq
U(t_f , t_i ) &=& 
\exp_{\leftarrow} \left(
-{i\over \hbar } \int^{t_f }_{t_i } dt {\cal H}(t) \right) .
\eeq
The operator $x(u)$ is $U^{\dagger}(u, t_{ini}) x U (u, t_{ini})$.
We take $t_{ini} \to -\infty$, and consider the deviation 
of the velocity $\dot{x}$ which has 
the form (\ref{deviation_of_v}). Then, we immediately obtain 
the response function, which is expressed using
the retarded Green function $G_{\dot{x}\, x}^{r} (t, u )$ as
\beq
\chi (t, u)= - \,G_{\dot{x}\, x}^{r} (t, u)=
{i\over \hbar}\Theta(t - u)\langle \left[
\dot{x} (t), x (u ) \right]\rangle . \label{retarded}
\eeq
Here, $\langle ... \rangle$ denotes an average over the initial state.
\section{Energy Dissipation and the Callen-Welton FDR}
We consider an energy dissipation flow from the system into the reservoir. 
By taking the derivative of the system's Hamiltonian with respect to time, we get the 
heat current operator
\beq
{\cal I} = -\sum_{\ell }
{\lambda_{\ell} \over m } p x_{\ell} 
+ \sum_{\ell} {\lambda_{\ell}^2 \over 2 m_{\ell } \omega_{\ell}^2 }
{1\over m} \left( xp + px \right)
, \label{current_operator}
\eeq
where positive current flows from the system into the reservoir.
To arrive at the average current at the steady state, we employ 
the technique of Keldysh green function. 
We use conventional notations for the Green functions, 
which are defined for arbitrary operators, $A$ and $B$, as
\beq
G_{A \, B}^{k}(t ,t' ) &=& 
-{i\over \hbar}
\left( \, \langle A (t )   B (t ' )  \rangle 
+         \langle B (t ' ) A (t )   \rangle \, \right)
, \nonumber \\
G_{A \, B}^{r,a}(t ,t' ) &=& 
-{i\over \hbar} \Theta (\pm (t - t'))
\langle [  B(t )  , A (t' ) ]\rangle  , \nonumber
\eeq 
where $G_{A \, B}^{k}(t ,t' ), G_{A \, B}^{r}(t ,t' )$, 
and $G_{A \, B}^{a}(t ,t' )$ are the Keldysh, retarded, and advanced 
Green function, respectively. 
We set the initial time to be $t_{ini}=-\tau/2$.
Using the green function, an average current is calculated as
\beq
I &=& \lim_{\tau\to\infty} {1\over \tau} \int_{-\tau/2}^{\tau /2} dt  
\langle {\cal I}' (t)  \rangle  + I_0 , \label{foemal1} \\
\langle {\cal I}' (t) 
\rangle  &=&
- {i \hbar \over 2}  \sum_{\ell}\lambda_{\ell}
{\partial \over \partial t_1 }G_{x, x_{\ell}}^{k} (t_1 , t_2 ) \Bigr|_{t_1 = t_2 = t} , ~~~~~~ \label{formal2}
\eeq
where $I_0 = \lim_{\tau\to\infty}{1\over \tau } \int_{-\tau /2}^{\tau /2}
dt \, {d \over dt }\langle x^2 (t) \rangle \int_{0}^{\infty }d\omega \gamma \, '  
(\omega ) /\pi $, which is the contribution from the second term in 
Eq.(\ref{current_operator}).
We assume that the initial density matrix of the system is an equilibrium 
distribution for the Hamiltonian without the 
potential part $V(x,t)$, i.e, ${p^2 / 2m } 
+ ({x^2 / 2}) \sum_{\ell} {\lambda_{\ell}^{2} 
/ ( m_{\ell} \omega_{\ell}^2} )$. 
This set up for initial state
enables us to use the Wick theorem to compute the Green functions. 
Perturbation expansions for contour-ordered Green functions 
are performed along the Schwinger-Keldysh contour 
depicted in Figure $1$. 
Using the Langreth rule for making the Keldysh Green function
in Eq.(\ref{formal2}) \cite{jauho}, we readily derive 
the average current 
\beq
I &=& I_0 +
\lim_{\tau\to\infty} {i \hbar \over 2 \tau}
\int_{-\tau/2}^{\tau /2} \int_{-\tau/2}^{\tau /2}
dt dt ' \nonumber \\
&\times& \Bigl\{ 
G_{\dot{x}\, x}^r (t,t') \Sigma^k (t' , t )
+ G_{\dot{x}\,x}^k (t,t' ) \Sigma^a (t' , t )
\Bigr\} .\label{formula1}
\eeq
In Eq.(\ref{formula1}), the function 
$\Sigma^k$ and $\Sigma^a$ are self-energy terms from the reservoirs, which 
calculated from the free green functions for reservoir's Hamiltonian. These
are written as 
\beq
\Sigma^k (t' ,t) \!\!&=&\!\!
{-2i\over \pi }\int_{0}^{\infty } \!\! d \omega \gamma \, ' (\omega ) 
{ \omega \over \tanh({\beta\hbar\omega \over 2}) } 
\cos (\omega (t' -t))
~,~~~ \nonumber \\
\Sigma^a (t' ,t) \!\!&=&\!\!
{-2\over \pi }\Theta (t - t' ){\partial \over \partial t'}
\int_{0}^{\infty }\!\! d \omega \gamma \, ' (\omega ) 
\cos (\omega (t' -t))
~.~~~ \nonumber
\eeq
\begin{figure}
\onefigure{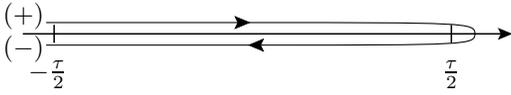}
\caption{Schwinger-Keldysh contour.}
\label{fig.1}
\end{figure}
The formula (\ref{formula1}) is 
valid for an arbitrary time-dependent driving field. 
In order to simplify it further, we consider the following quantity 
for the Green function
\beq
G^{\alpha}_{AB} ( \omega ) = \lim_{\tau\to\infty}
{1\over \tau }
\int_{-\tau/2}^{\tau/2 } \!
\int_{-\tau/2}^{\tau/2 } \! d t d t'  G^{\alpha}_{AB} (t  , t' )
e^{i \omega  t }e^{ - i \omega  t '}  , \nonumber
\eeq
where $\alpha$ represents $r,a$, and $k$. 
When translational invariance in time
is satisfied, this reduces to the usual expression
of the Fourier transformation. 
Let $\chi\,'(\omega)$ denote
$-{\rm Re}[G_{\dot{x} x}^{r} (\omega )]$.
Then, a straightforward modification leads to
\beq
I &=&{1\over \pi } 
\int_{0}^{\infty } \! d \omega  \gamma \,' (\omega )
\Bigl[ 2\pi v_s^2 \delta (\omega ) +  C (\omega ) -
{  \chi\,' (\omega )  \hbar \omega \over 
\tanh({\beta \hbar\omega \over 2}) }\Bigr]
] , 
\nonumber \\
\label{result1}
\eeq
where we used the integral by parts to remove the term $I_0$.
Here, $C(\omega )= (i\hbar/2)
G_{\dot{x} \dot{x}}^{k} (\omega )  - 2 \pi v_s^2 \delta (\omega )$, and
$v_s$ is the average velocity.
In an equilibrium state where no net energy dissipation flow exists, the Callen-Welton
FDR, 
$ C(\omega ) =  
\chi\,' (\omega ) \hbar \omega /\tanh(\beta\hbar \omega / 2 ) $ 
is satisfied \cite{callen_welton}. In the limit $\hbar\to0$ and 
by inserting $\gamma \,' (\omega) = \gamma$, 
the above equation reproduces Eq.(\ref{HS}). 

\section{Concluding Remarks}
Several general principles exist for nonequilibrium phenomena. 
In the linear response regime, the validity of
the Onsager reciprocity and the Green-Kubo relations have been 
established.
The fluctuation theorems and Jarzynski equality are being investigated
in numerous models and experiments, and 
they seem to be the exact relations valid arbitrarily far from equilibrium.
Eqs.(\ref{HS}) and (\ref{result1}) are 
also interesting, showing that the degree of FDR violation
is related to energy dissipation in Langevin dynamics. 
This would be important, because Langevin dynamics is ubiquitous 
in realistic systems.
Quantum Langevin dynamics is believed to have 
wide applicability in many realistic systems, including
metallic tunnel junctions with capacitances 
and superconducting junctions \cite{weiss,CL,ambegaokar1,exp}. 
In those cases, the system and reservoir variables 
represent the variables in electrical circuits, 
and the external force is realized by an electric current.
Circuit-realization 
of a driven harmonic trapped particle was proposed in Ref.\cite{zon04}.
The quantum case for this would be an interesting relevant system.

In the present study, Eq.(\ref{formula1}) is the key equality for
deriving the result (\ref{result1}). To derive Eq.(\ref{formula1}), 
linear coupling between the system and thermal environment was critical.
Although linear couplings should be the dominant contributions 
in most realistic systems, nonlinear couplings, where a nonlinear function 
of $x$ couples with the bath variables, are also possible \cite{tanimura}. 
It is possible to generalize Eq.(\ref{formula1}) to 
nonlinear coupling cases. In general,
when we change the coupling form in the Hamiltonian (\ref{ham1}) 
as $\lambda_{\ell} x_{\ell}
x \to \lambda_{\ell} x_{\ell} f(x)$, 
a different expression of average current from Eq.(\ref{result1}) is obtained as
\beq
I \!&=&\!{1\over \pi } 
\int_{0}^{\infty } \! d \omega  \gamma \,' (\omega )
\Bigl[ {i\hbar \over 2} G_{\dot{f} \, \dot{f} }^k ( \omega ) 
+{\hbar \omega  {\rm Re} [G_{\dot{f} \, f }^r (\omega ) ]
\over 
\tanh({\beta \hbar\omega \over 2}) }
\Bigr]
. ~~~~~~~
\eeq
This means that the expression depends on types of coupling form.
It would be important to figure out how Eqs. (\ref{HS}) and 
(\ref{result1}) are generalized,
if we consider other types of reservoirs and dynamics.
Energy dissipation of spin dynamics would be an 
important problem to be studied.

Another intriguing problem might be on higher order fluctuations of
energy dissipation flow. Nonequilibrium fluctuations increases in time
unlike equilibrium ones. To study characteristics of fluctuations, 
it is convenient to use the technique of counting statistics 
\cite{levitov_lesovik,QNoise}, which is equivalent to 
the protocol for obtaining {\em fluctuation theorem} in quantum systems
\cite{saito_dhar,tasaki_matsui,kurchan} and work distribution \cite{hanggi}. 
It is possible to
derive an explicit form of characteristic function, which generates
not only average current but also any orders 
of cumulants of dissipation flow. It can reproduce the present result 
(\ref{result1}). Systematic derivatives of the characteristic function 
with respect to a counting field generates any orders of fluctuations. 
We hope that this study encourages further studies on 
energy dissipation at far-from-equilibrium conditions 
in quantum systems.

\acknowledgments
The author would like to thank S. Sasa, Y. Utsumi, T. Kato, and S. Tasaki 
for useful comments and discussions. 
This work was supported by the Grant in Aid from 
the Ministry of Education, Sports, Culture and Technology of 
Japan (No.~19740232).

\end{document}